\documentclass[aps,preprint,prb,showpacs,preprintnumbers,amsmath,amssymb]{revtex4}

\usepackage{graphicx}
\usepackage{bm}

\begin{document}

\title{Probing the Shape of Quantum Dots \\ with Magnetic Fields}

\author{P.\,S. Drouvelis}%
 \email{panos@tc.pci.uni-heidelberg.de}
\affiliation{%
Theoretische Chemie, Universit\"at Heidelberg,
Im Neuenheimer Feld 229, D-69120 Heidelberg, Germany}%

\author{P. Schmelcher}%
 \email{peter@tc.pci.uni-heidelberg.de}
\affiliation{%
Theoretische Chemie, Universit\"at Heidelberg,
Im Neuenheimer Feld 229, D-69120 Heidelberg, Germany}%
\affiliation{%
Physikalisches Institut, Philosophenweg 12, Universit\"at Heidelberg,
D-69120 Heidelberg, Germany}%

\author{F.\,K. Diakonos}%
 \email{fdiakono@cc.uoa.gr}
\affiliation{%
Department of Physics, University of Athens, GR-15771 Athens,
Greece}%

\date{\today}

\begin{abstract}
A tool for the identification of the shape of quantum dots is developed. By preparing a two-electron quantum dot,
the response of the low-lying excited states to a homogeneous magnetic field, i.e. their spin and parity oscillations, is studied
for a large variety of dot shapes. For any geometric configuration of the confinement we encounter characteristic
spin singlet - triplet crossovers. The magnetization is shown to be a complementary tool for probing the shape of the dot.
\end{abstract}
\maketitle

\section{INTRODUCTION}

The electronic structure of few-electron quantum dots (QDs) has been probed experimentally
and provided us with new rich physics \cite{Jacak97,Reimann02,Tarucha96,Kouwenhoven01}.
The basic physical picture consists of electrons effectively trapped by an in-plane parabolic artificial potential,
a model which is compatible with the insensibility of the QDs to far-infrared radiation and is reflected in theory
by the generalized Kohn theorem \cite{Kohn61, Peeters90}. Furthermore, for circular symmetric confinement
a wide range of experimentally probed phenomena, such as addition energy spectra \cite{Tarucha96,Kouwenhoven01}
or inelastic light scattering \cite{Brocke03} measurements, could be explained
by adopting such a model. The response of QDs to homogeneous magnetic fields has triggered rich scenarios.
Among others, these are the magic angular momentum numbers, the configurations and the oscillations
of the total spin \cite{Maksym92,Maksym93} as well as the parity oscillations for the ground state of
a QD \cite{Wagner92,Dineykhan99} with changing field strength. The electron-electron interaction
plays hereby a crucial role.

Introduction of the anisotropy breaks the rotational symmetry thereby altering the dot's electronic and dynamical properties.
The shell structure in the addition energy spectra \cite{Reimann02,Ezaki97,Austing99} and the level clustering
in the two-electron QD excitation spectrum \cite{Drouvelis03a,Drouvelis03b} are eliminated and
dynamically the anisotropy serves as a rapid path to chaos \cite{Drouvelis03a,Drouvelis03b}.
The spin configurations obey the Hund's rule for small QDs while for larger ones they change to
a more Pauli-like behavior \cite{Fujito96,Hirose99}. In three dimensions the anisotropy has been introduced
with a magnetic field which therefore controls the symmetries \cite{Simonovic03}.
However, despite the enviable advances in revealing \cite{Kouwenhoven01} and explaining \cite{Reimann02}
the electronic structure of parabolic QDs, the question for the exact shape of a dot,
as it is realized experimentally, can not be answered precisely. A major approximation often made, suggests
an analogy between the geometric configuration of the device and the effective confinement potential
\cite{Reimann02,Austing99}. However, it is being admitted that it is nearly impossible to
derive the exact theoretical model of the confinement from the geometry of the setup of the device.
Even within circular geometry the precise electronic properties depend on the details of the device \cite{Reimann02}.

In the present article the properties of two-electron parabolic QDs in magnetic fields
are investigated for the full deformation regime from circularly shaped to wirelike dots.
In Sec. II we provide the Hamiltonian within the framework of the effective mass approximation
and describe its general symmetries as well as our computational approach. 
In Sec. III we present and discuss our results. In particular, the properties of the two-electron anisotropic
QD (Sec. III.A) and the isotropic two-electron QD in magnetic field (Sec. III.B) are summarized in order to
become acquainted with the indivudual effects of the anisotropy and magnetic field in our two-electron
system. Sec. III.C provides the response of the spin and parity symmetries of the ground and few low-lying states as well
as the magnetization to the magnetic field for the full deformation regime addressed here.
We encounter unique features for every dot shape. Ultimately, we develop a tool for the diagnosis
of the shape of dots, as they are realized experimentally.

\section{HAMILTONIAN AND COMPUTATIONAL APPROACH}

In the effective mass approximation the conduction band electrons confined in a two-dimensional general
parabolic quantum dot in a magnetic field {\bf{B}} $ = (0,0,B)$  are described by the Hamiltonian
${\cal{H}}={\cal{H}}_{CM} + {\cal{H}}_{r}$ with

\begin{eqnarray}
{\cal{H}}_{CM} &=& \frac{1}{4m_e} ({\bf{P}} + 2e{\bf{A(R)}})^2 + m_e \omega_o^2 \left( cos^2\phi~X^2 + sin^2\phi~Y^2 \right) \\ \label{eq1}
{\cal{H}}_{r} &=& \frac{1}{m_e} ({\bf{p}} + \frac{e}{2}{\bf{A(r)}})^2 + \frac{m_e}{4} \omega_o^2 \left(cos^2\phi~x^2 + sin^2\phi~y^2 \right)
+ \frac{e^2} {4\pi\epsilon\epsilon_o~|{\bf{r}}|} 
\label{eq2}
\end{eqnarray}

\noindent where we choose for the vector potential the symmetric gauge {\bf{A(r)}} = $\frac{1} {2}$ ({\bf{B}} $\times$ {\bf{r}})
and $e, m_e, \epsilon, \omega_o, \phi$ are the electron charge, effective mass, dielectric constant, the characteristic frequency
and the anisotropy parameter, respectively. Small and capital letters refer to the relative and center of mass
degrees of freedom, respectively. In the following we adopt the typical
values for a GaAs dot (effective Bohr radius $a_B^*=9.8$ nm, effective Hartree $Ha^*=11.8$ meV, $\hbar \omega_o=4.96$ meV
and 1 effective unit (e.u.) of field strength corresponds to $6.925$ Tesla).
Quantization of ${\cal{H}}_{CM}$ is straightforward. Direct observation of the
electronic properties due to ${\cal{H}}_{r}$ via far infrared spectroscopy is prohibited, since radiation in the dipole approximation
contains only CM degrees of freedom. In the following we focus on the non-trivial relative motion ${\cal{H}}_{r}$. 
Parity (${\bf{r}} \rightarrow -{\bf{r}}$) and spin are interrelated symmetries due
to the Pauli exclusion principle and we
encounter spin singlet eigenfunctions with even spatial symmetry $\Psi({\bf{r}})=\Psi(-{\bf{r}})$ and spin triplet
eigenfunctions with odd spatial symmetry $\Psi({\bf{r}})=-\Psi(-{\bf{r}})$.
The two characteristic frequencies of the confinement are $\omega_x = \omega_o~cos \phi$ and
$\omega_y = \omega_o~sin \phi$. For $\phi=45^{\circ}$ the dot has a circular shape.
With increasing $\phi$ it obtains an elliptic shape and ends in a wirelike dot
for $\phi \rightarrow 90^{\circ}$ ($\omega_x \rightarrow 0$, $\omega_y \rightarrow \omega_o$).

To investigate the two-electron QD we solve the corresponding Schr\"odinger equation
using a full configuration interaction approach with the harmonic oscillator-like basis set
\begin{equation}
\Phi_{n_xn_y} = A(n_x,n_y) H_{n_x}(\sqrt{c_1}~x)H_{n_y}(\sqrt{c_3}~y) \\
\times e^{- \frac{c_1}{2}x^2 - \frac{c_3}{2}y^2 + i(\lambda - \frac{c_2}{2})xy}
\label{eq3}
\end{equation}

\noindent which leads to an algebraic eigenvalue problem. In Eq. (\ref{eq3}) $A(n_x,n_y)$ is the normalization constant,
$c_1 = M_1 \omega_1/c$, $c_3 = M_2 \omega_2/c$, $c_2 = 2 \mu M_1 \omega_1 M_2 \omega_2 / c$,
$c = \mu^2 M_1 M_2 \omega_1 \omega_2 +1$, $\mu = -2L/(m_e\omega_0p)$, $\lambda = [m_e \omega_0 L (2+L^2)]/[4(cos(2\phi) - p)]$ 
, $M_{1,2} = m_ep/(p-cos(2\phi) \mp L^2)$, $\omega_{1,2} = (\omega_0 / \sqrt{2}) \sqrt{1+L^2 \mp p}$, $L = eB / m_e \omega_0$
and $p = \sqrt{(1+L^2)^2 - sin^2(2 \phi)}$. All the units are scaled appropriately. The eigenfunctions $\Psi_{n_1,n_2}$ of the charged
anisotropic harmonic oscillator (corresponding to ${\cal{H}}_r$ without Coulomb repulsion)
can be obtained analytically \cite{Schmelcher94}. The exponential function of our basis set (\ref{eq3}) represents exactly
the exponential part contained in $\Psi_{n_1,n_2}$. The Hermite polynomials contained in $\Psi_{n_1,n_2}$ 
can be obtained through a {\it{finite}} superposition of the corresponding
Hermite polynomials in the orbitals $\Phi_{n_x,n_y}$. In order to arrive at the eigenstate $\Psi_{n_1,n_2}$ we need to superimpose
Hermite polynomials of degree $n_x = n_y = n_1 + n_2$ for both $x$ and $y$ directions. For the evaluation
of the Hamiltonian matrix elements (without the Coulomb interaction) we subtracted from Eq. (\ref{eq2}) the Hamiltonian
whose eigenfunctions are $\Phi_{n_x,n_y}$. The remaining electron-electron integral was evaluated numerically.
The Coulomb repulsion term was replaced by an auxiliary Gaussian integral
and the resulting three-dimensional integral was evaluated by combining Gauss-Hermite exact quadratures with respect to
$x$ and $y$ directions and a Gauss-Kronrod quadrature in the remaining auxiliary variable. For details we refer the
reader to Refs. \onlinecite{Drouvelis03a,Drouvelis03b}.

\section{RESULTS AND DISCUSSION}

\subsection{No magnetic field}

Before we proceed with the analysis of our results let us briefly discuss some principal properties of our system without magnetic field
illustrated in Fig. 1 (see also Refs. \onlinecite{Drouvelis03a,Drouvelis03b}).
For $\phi=45^\circ$, ${\cal{H}}_{r}$ is rotationally symmetric and $L_z$ is conserved. The sequence
of excited states with increasing energy is: ($m;S$) = ($0;0$),($\pm1;1$),
($\pm2;0$),($0;0$),($\pm3;1$),($\pm1;1$),$\ldots$, where $m$ and $S$ are
the magnetic quantum number and the total spin, respectively.
Introduction of the anisotropy splits the doubly degenerate levels (for $m\neq0$) and leads to spin singlet - spin triplet (ST) crossings.
At $\omega_y:\omega_x = 2:1$ the system becomes integrable due to the non-linear
constant of motion 

\begin{eqnarray}
{\Lambda}=\{{L}_z,{p}_x\} + \frac{\omega_x^2}{2}yx^2
 - \frac{y}{\sqrt{x^2+y^2}}
\label{eq4}
\end{eqnarray}

The operator $\Lambda$ commutes with the $x$-parity and anticommutes with the $y$-parity. These symmetries
result in ST degeneracies which occur systematically in the excitation spectrum (see Fig. 1).
For $\phi \rightarrow 90^\circ$ the low-lying states arrange themselves in energetically well-separated pairs involving
one spin singlet and one spin triplet state.

\begin{figure}
\centering
\includegraphics[width=6cm,angle=-90]{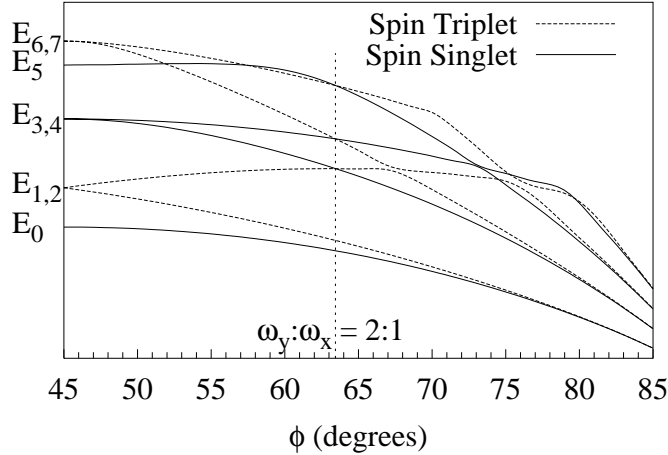}
\caption{Schematic diagram of the energy of the ground and first seven excited states for $B=0$ with increasing $\phi$. The vertical
line indicates the angle for which the system is integrable.}
\label{figure1}
\end{figure}

\subsection{Isotropic case in a magnetic field}

The two-electron isotropic QD in a magnetic field has been well studied \cite{Wagner92,Merkt91,Pfannkuche93a,Pfannkuche93b} and particular
analytical solutions have been found \cite{Taut94}. The system is integrable, with $L_z$ being a constant of motion.
A major feature of this system are the parity oscillations of the ground state \cite{Wagner92}, being a clear imprint
of the Coulomb interaction. By increasing the magnetic field, confinement strengthens thereby increasing the Coulomb
energy between the two electrons. Consequently, the system attempts to establish itself in a lower energy
by pushing apart the electrons into states with higher $|m|$. Hence, by increasing the field the ground state
pumps itself into an energetically favourable state, thereby flipping the spin, with the following sequence:
($m;S$) = ($0;0$),($-1;1$),($-2;0$),($-3;1$),$\ldots$, i.e. we encounter ST oscillations for the ground state.
For excited states an increasing number of crossings occur \cite{Merkt91} due to the participation
of states covering an increasing range of $m$. For increasing field strength, the level spacing $\Delta E_i = E_{i+1} - E_i$
({\it{i}} determines the degree of excitation and takes even values $i = 0,2,4,...$ within our study)
oscillates between zero and a maximum value $\Delta E_{max}$.
For stronger fields the amplitude $\Delta E_{max}$ reduces and the slope
of the oscillating energy curves converges to a constant value. This fact leaves its fingerprints
in the magnetization which for zero temperature is defined by $M(B) = -(\frac{\partial E_0}{\partial B})$
with $E_0(B)$ being the ground state energy. The ST crossovers will then appear as steps in the magnetization.
With increasing temperature these steps smear out and result in a smooth diamagnetic behavior.
Typically, for $T<0.1K$ the magnetization's signal reveals the ST oscillations.

\subsection{Anisotropic configuration in the magnetic field}

\begin{figure}
\centering
\includegraphics[width=14cm]{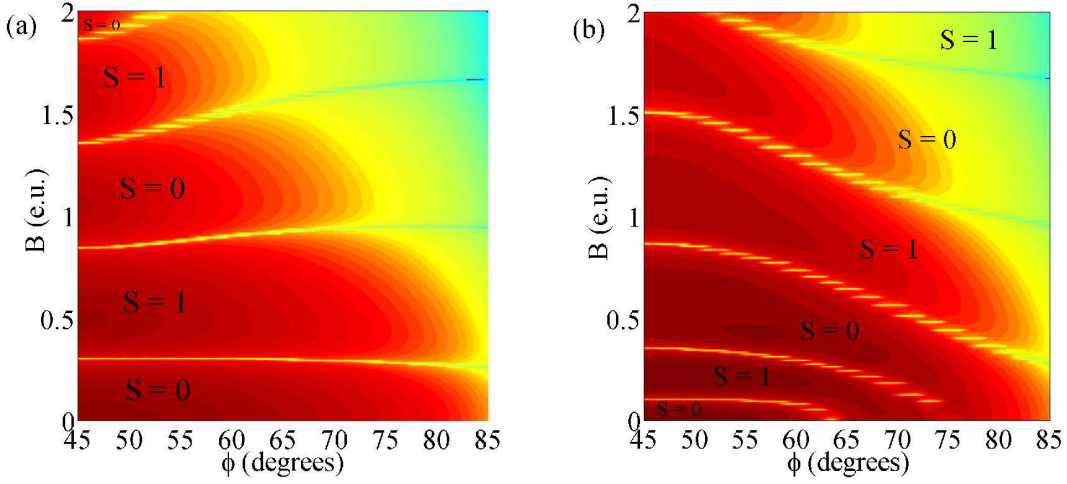}
\caption{(Color online) Domains of spin multiplicity in the ($B,\phi$) plane for (a) the ground state and (b) the fourth excited state.
Color indicates the energy difference $\Delta E_0$ and $\Delta E_4$, respectively, on a logarithmic scale.
Deep red (dark regions) and yellow to green (bright regions) correspond to large and small spacings, respectively.
The yellow (bright) curves form the borders between the different ST-symmetry domains.}
\label{figure2}
\end{figure}

Let us now proceed with the introduction of the anisotropy in the two-electron isotropic QD in the magnetic field.
The rotational symmetry breaks and a large number of avoided crossings between states with identical symmetry occur.
Fig. 2(a) shows the spin multiplicity $S=0,1$ of the ground state in the ($B,\phi$) plane. 
We observe alternating islands of different total spin. For $\phi = 45^\circ$ 
and the given range of magnetic field we receive ST
oscillations in correspondence with Ref. \onlinecite{Wagner92}. Introducing the anisotropy, for
$B=0$, splits the first excited doubly degenerate triplet state ($E_{1,2}$). As a result $\Delta E_0$ decreases
gradually with $\phi$ and in the wirelike limit $\phi \rightarrow 90^\circ$, $E_1$
approaches $E_0$ ($\Delta E_0 \rightarrow 0$) forming an energetically well-separated ST pair (see Fig. 1). For fields $B \lesssim 1$
the border curves separating the domains of different spin multiplicity depend very weakly on the field strength.
For higher field strengths $1<B<2$ a more significant change can be observed
as e.g. the third ($S=0$) and fourth ($S=1$) domain slightly spread and higher domains such as the fifth ($S=0$) domain for
$\phi>54^\circ$ (see upper left corner of Fig. 2(a)), are suppressed.

The third and fourth excited state for $\phi = 45^\circ$ and $B=0$ form a doubly degenerate spin singlet pair ($E_{3,4}$). By introducing the
anisotropy the corresponding energy splits and at $\omega_y:\omega_x=2:1$ the energy $E_3$ becomes degenerate with the
energy $E_2$ of the lower excited spin triplet state, an effect of the constant of motion
and its associated symmetries for this particular anisotropic configuration
(see Sec. III.A). Hence, for the second excited state (corresponding to $E_2$),
we observe a ST transition already in the absence of the field,
being a pure consequence of the geometric change of the confinement.
For stronger anisotropies ($\frac{\omega_y}{\omega_x}>2$) the energies $E_2,E_3$ of the two states
converge again and form an energetically well-separated pair of energies. By introducing
the magnetic field, the spin-symmetry domains change more significantly compared to the ground state. Hence, for $E_2$,
the first ($S=1$) domain, with increasing $\phi$, will be suppressed at $\omega_y:\omega_x=2:1$ by the following
($S=0$) domain. The domains, for stronger fields, slightly widen after the suppression of the first $S=1$ domain.

Fig. 2(b) shows the domains of spin multiplicity $S=0,1$ in the ($B,\phi$) plane for the fourth excited state.
The fourth excited state ($E_4$) is a spin singlet,
for $B=0$ and $\phi=45^\circ$, and at $\omega_y:\omega_x=2:1$ ($\phi \approx 63.4^\circ$)
it crosses with the energy curve $E_5$ (see Figs. 1,2(b)) thereby becoming a spin triplet.
For stronger anisotropies and due to the large number of excited states,
an 'accidental' crossing with a spin singlet state occurs and the eigenstate belonging to $E_4$ restores its initial
parity and spin-multiplicity. In the wirelike limit, a third pair of energetically well-separated states is formed.
For weak magnetic fields and changing $\phi$, the first ($S=0$) domain disappears at $\omega_y:\omega_x \approx 2:1$ and a
($S=1$) domain follows. For $\phi \gtrsim 75^\circ$ the $S=1$ domain is suppressed, due to
the above-mentioned 'accidental' crossing, by a second $S=0$ domain. For stronger fields, the domains
evolve smoothly and slightly widen after the accidental crossing at $\phi \approx 75^\circ$.
From Fig. 2(b) it can be seen that the border lines separating the spin domains in the
($B,\phi$) plane strongly depend on the anisotropy $\phi$.

{\it{We therefore conclude that the map of ST domains in the ($B,\phi$) plane for low-lying
states shows unique features that are characteristic for the shape of the dot.}}

\begin{figure}
\centering
\includegraphics[width=6.5cm,angle=-90]{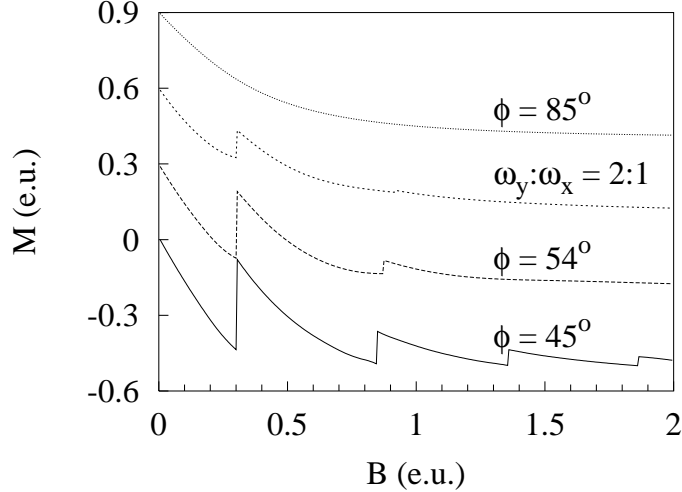}
\caption{The magnetization $M(B)$ for various anisotropies $\phi$. The different curves are offset
by 0.3.}
\label{figure3}
\end{figure}

An important tool to study the response of the QD to the presence of the external magnetic field
is its magnetization $M$. Calculations within the Hartree approximation for three and four electrons have shown that
$M$ is sensitive to the number of electrons and the shape of the dot \cite{Magnusdottir00}.
Fig. 3 shows $M(B)$ for the two-electron QD and various anisotropies. For $\phi = 45^\circ$ the curve shows four steps whose positions
correspond to the ground state ST crossovers. For lower magnetic fields the steps are more pronounced due to the
very different slopes of the crossing energy curves. Hence, the magnetization contains also information
on $\Delta E_0(B)$. By introducing the anisotropy, $\Delta E_0(B=0,\phi)$ decreases and the amplitude of the
oscillations $\Delta E_{max}$ also decreases thereby reducing the height of the steps.
Consequently, the steps with increasing $\phi$ gradually vanish and finally, we obtain a smooth
curve for $\phi = 85^\circ$. {\it{The magnetization, therefore, can reveal also significant information on the shape of the dot,
requiring the study of the ground state only.}}

The above discussion holds if the contribution of the Zeeman term $E_S(B) = g^* \mu_BBS_z$, (with $\mu_B$ being the Bohr magneton
and $g^* = -0.44$ for GaAs) is neglected. $E_S$ splits the threefold degeneracy of the spin triplet states while the
spin singlet states remain unchanged.
Hence, we expect a suppression of the spin singlet ground states for field strengths where $\Delta E_0(B)$ is
smaller than $E_S(B)$. As a result, for high magnetic fields, $E_S$ dominates and the ground state is always a spin triplet
\cite{Wagner92}. Fig. 4 shows the spin multiplicity of the ground state in the ($B$,$\phi$) plane in the
presence of the Zeeman term. For $\phi=45^\circ$ the ground state shows ST oscillations. Starting from
$B=0$ and increasing $B$, the interval of field strengths of the first $S=0$ domain is approximately the same as for $g^* = 0$.
However, the following $S=1$ domains are enlarged and the second $S=0$ domain shrinks significantly.
Introduction of the anisotropy causes a reduction of $\Delta E_0$ which is not noticeable for low field strengths
but leads, for stronger fields, to the disappearance of the second spin singlet island
already for $\phi \gtrsim 48^\circ$. Hence, apart from this spin singlet island, the ground state
is given by a spin triplet for $B > 0.3$, in the regime from circular to intermediate anisotropies ($\phi \lesssim 65^\circ$).
For higher anisotropies, $\Delta E_0$ decreases significantly, resulting in a smooth elimination
of the first spin singlet domain as $\phi$ approaches the wirelike limit.

\begin{figure}
\centering
\includegraphics[width=7cm,angle=-90]{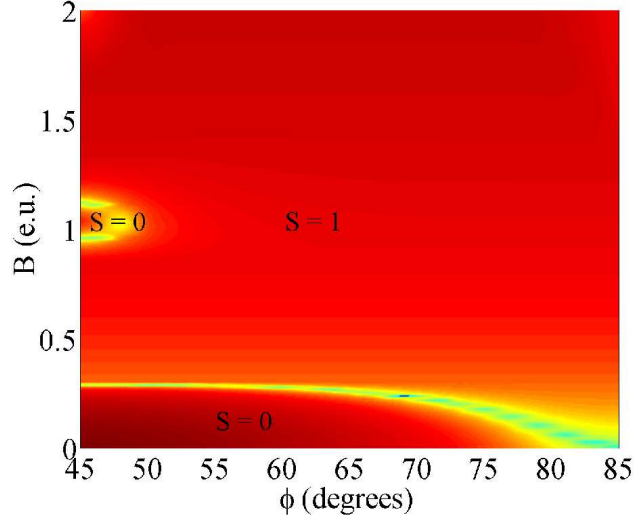}
\caption{(Color online) Domains of spin multiplicity of the ground state in the ($B$,$\phi$) plane
at the presence of the Zeeman spin splitting.}
\label{figure4}
\end{figure}

An important issue to be discussed is the effect of the Zeeman splitting for higher excited states
in the presence of anisotropy. Clearly, for larger fields, the ST oscillations
will be suppressed in favour of the spin triplet symmetry due to the dominance
of the Zeeman splitting. For lower fields, however, the ST oscillations persist with varying $\phi$.
This statement holds typically for $B \lesssim 1$ and a significant number of excitations.
{\it{The results obtained above for $g^*=0$ therefore transfer immediately to the case $g^* = -0.44$.}}

\section{CONCLUSIONS}

To conclude, a tool for the identification of the shape of a QD has been developed, by preparing a two-electron
QD and studying its response, when switching on a homogeneous magnetic field.
We examined low-lying excited states, that are accessible experimentally,
with e.g. tunneling spectroscopy. By applying a bias voltage between the
two reservoirs weakly coupled to the dot, the Coulomb blockade region can become narrower and the
extra tunneling channels, corresponding to the excited states, can be detected as an increase in the current \cite{Kouwenhoven01}.
The parity and spin symmetries of these states change uniquely with varying the shape $\phi$, for field strengths
up to several Tesla. Hence, the detection of the positions of the ST crossovers for low-lying excited states
allows us to conclude uniquely on the shape $\phi$ of the QD. As an alternative, independent possibility
to confirm the results we suggest magnetization measurements. Our investigation yields the $\phi$-dependence of
the magnetization showing a gradual transition from a step-like to smooth behavior.
In the presence of the Zeeman term ($g^* = -0.44$) and for weak to intermediate field strengths
our results remain practically unchanged and can be equally applied.
In the strongly anisotropic regime, already the ground state ST crossover curve 
in the ($B$,$\phi$) plane shows a monotonic change with $\phi$.
The method developed here applies, therefore, to the full deformation regime.

\begin{acknowledgments}
We thank O. Alon for illuminating discussions.
Financial support in the framework of the IKYDA program of the DAAD
(Germany) and IKY (Greece) is gratefully acknowledged.
\end{acknowledgments}

\end{document}